\documentclass[a4paper,12pt]{article}
\usepackage{amssymb,amsmath}
\usepackage{a4wide}
\usepackage{epsfig}

\newcommand{\RR}{{\mathbb{R}}}

\newcommand{\ZZ}{{\mathbb{Z}}}
\newcommand{\ii}{{\mathrm{i}}}
\newcommand{\ee}{{\mathrm{e}}}

\newcommand{\st}{\tilde{s}}
\newcommand{\half}{{\scriptstyle\frac{1}{2}}}
\newcommand{\quar}{{\scriptstyle\frac{1}{4}}}
\newcommand{\thrquar}{{\scriptstyle\frac{3}{4}}}
\newcommand{\tr}{\mathop{\mathrm{tr}}\nolimits}
\renewcommand{\Re}{\mathop{\mathrm{Re}}\nolimits}
\renewcommand{\Im}{\mathop{\mathrm{Im}}\nolimits}
\newcommand{\cM}{{\cal M}}
\newcommand{\cE}{{\cal E}}
\newcommand{\cS}{{\cal S}}
\newcommand{\cP}{{\cal P}}
\newcommand{\vR}{{\bf R}}
\newcommand{\ve}{\varepsilon}

\title{Dynamics of Monopole Walls}
\author{R.\ Maldonado\footnote{email address: rafael.maldonado@durham.ac.uk}
  \,\, and R.\ S.\ Ward\footnote{email address: richard.ward@durham.ac.uk}
  \bigskip
  \\Department of Mathematical Sciences,
  \\Durham University, Durham DH1 3LE.}
\date{\today}

\begin{document}



\maketitle

\begin{abstract}
\noindent The moduli space of centred Bogomolny-Prasad-Sommmerfield
2-monopole fields is a 4-dimensional manifold $\cM$ with a natural metric,
and the geodesics on $\cM$ correspond to slow-motion monopole dynamics.
The best-known case is that of monopoles on $\RR^3$, where $\cM$ is the
Atiyah-Hitchin space. More recently, the case of monopoles periodic in one
direction (monopole chains) was studied a few years ago. Our aim in this
note is to investigate $\cM$ for doubly-periodic fields, which may be
visualized as monopole walls. We identify some of the geodesics on $\cM$
as fixed-point sets of discrete symmetries, and interpret these in terms of
monopole scattering and bound orbits, concentrating on novel features
that arise as a consequence of the periodicity.
\end{abstract}



\section{Introduction}
The observation that the dynamics of Bogomolny-Prasad-Sommmerfield (BPS )
monopoles can be approximated as geodesics on the moduli space $\cM$
of static solutions \cite{M82} has proved to be far-reaching. Not only does it
reveal much about monopole dynamics, but the moduli spaces themselves
are of considerable interest, for example in string theory.
The best-known case is that of the centred 2-monopole system on $\RR^3$,
where $\cM$ is a 4-dimensional asymptotically-locally-flat (ALF) space,
namely the Atiyah-Hitchin manifold \cite{AH88, MS04}. For monopoles
periodic in one direction, in other words on $\RR^2\times S^1$, the
asymptotic behaviour of the centred 2-monopole moduli space is
different, and is called ALG \cite{CK02}. In this case, the generalized Nahm
transform has been used to describe some of the geodesics on the
moduli space, and their interpretation in terms of periodic monopole dynamics
\cite{HW09, MW13}.

This paper focuses on the doubly-periodic case, namely BPS monopoles
on $T^2\times\RR$, also referred to as monopole walls
\cite{W07, CW12}. An $N$-monopole field which is periodic in the $x$-
and $y$-directions may be viewed as a set of $N$ monopole walls,
each extended in the $xy$-direction. Much is known about the general
classification of the moduli spaces of such solutions,
and their string-theoretic interpretation \cite{CW12, C14}.
We shall restrict our attention here to the case of smooth 2-monopole fields
with gauge group SU(2); the centred moduli space $\cM$ is then a
four-dimensional hyperk\"ahler manifold with so-called ALH boundary
behaviour \cite{C11}.  The asymptotic form of its metric has recently been
derived \cite{HKM14}. Our aim here is to identify some of the geodesics on $\cM$
as fixed-point sets of discrete symmetries, and to interpret these in terms of
monopole scattering, concentrating on novel features that arise as a
consequence of the periodicity.

The system, therefore, consists of a smooth SU(2) gauge potential $A_j$
on $T^2\times\RR$, plus a Higgs field $\Phi$ in the adjoint representation.
The fields satisfy the Bogomolny equation $D_j\Phi=-B_j$, where
$B_j=\half\ve_{jkl}F_{kl}$ is the SU(2) magnetic field. The coordinates
are $x^j=(x,y,z)$, where $x$ and $y$ are periodic with period~$1$,
and $z\in\RR$. The boundary condition (see \cite{W07, CW12} for more
detail) is $|\Phi|/|z|\to{\rm const}$ as $z\to\pm\infty$. There are two
topological charges $Q_{\pm}$, which are non-negative integers defined
in terms of the winding number of $\Phi$. More precisely, if
$\Phi_c:=\Phi|_{z=c}$, then $\hat{\Phi}_c:=\Phi_c/|\Phi_c|$ is a map from
$T^2$ to $S^2$, and we define $Q_{\pm}:=\pm\deg\hat{\Phi}_{\pm c}$
for $c\gg1$.
The number of monopoles is $N=Q_++Q_-$, and we are interested
in the case $N=2$, so there are three possibilities, namely
$(Q_-,Q_+)=(1,1)$, $(0,2)$ or $(2,0)$. In fact, the corresponding moduli
spaces are isometric \cite{C14}. In what follows, we shall concentrate
on the $(1,1)$ wall, namely $Q_-=Q_+=1$.


\section{Parameters and moduli of the $(1,1)$ wall}
We begin by reviewing the parameters, the moduli, the
energy, and the spectral data of the $(1,1)$ wall, using the same
conventions and notation as in \cite{CW12}. There exists a (non-periodic)
gauge such that the boundary behaviour of the fields is
\begin{equation} \label{AsympField}
  \Phi\sim2\pi\ii(z+M_{\pm})\sigma_3, \quad
              A_j\to\pi\ii(y-2p_{\pm}, -x-2q_{\pm},0)\sigma_3
\end{equation}
as $z\to\pm\infty$. The six real constants $(M_{\pm},p_{\pm},q_{\pm})$
are the boundary-value parameters, with $M_{\pm}\in\RR$ and
$p_{\pm},q_{\pm}\in(-\half,\half]$. Fixing the centre-of-mass of the system
amounts to fixing $(M_-,p_-,q_-)$ in terms of the other three parameters
$(M_+,p_+,q_+)$. Henceforth, we fix the centre-of-mass to be at the
point $(x,y,z)=(\half,\half,0)$, and the field is then invariant (up to a
gauge transformation) under the map
$(x,y,z)\mapsto(1-x,1-y,-z)$ plus $\Phi\mapsto-\Phi$.
In effect, the system as a whole has infinite mass,
and only the relative separation and phase of the two monopoles
appear in the moduli space; the space of fields with fixed
$(M_{\pm},p_{\pm},q_{\pm})$, modulo gauge transformations,
is our four-dimensional moduli space $\cM$.

The energy density is $\cE = |D\Phi|^2+|B|^2$, and $\cE\to8\pi^2$
as $z\to\pm\infty$. The total energy, {\sl ie.}\ $\cE$ integrated over
$T^2\times\RR$, is consequently infinite. But the cut-off energy
\begin{equation} \label{Energy}
  E_L = \int_{-L}^L dz\,\int\left(|D\Phi|^2+|B|^2\right)\,dx\,dy
\end{equation}
is finite, and if $L\gg-M_+$ it equals the Bogomolny bound \cite{W07}
\begin{equation} \label{Bog}
  E_L=16\pi^2(L+M_+).
\end{equation}
Spectral data for this system may be defined as follows \cite{CW12}. Put
\[
  W_x=\tr\cP\exp\int_0^1(-A_x-\ii\Phi)\,dx, \quad
 W_y=\tr\cP\exp\int_0^1(-A_y-\ii\Phi)\,dy.
\]
Then $W_x$ and $W_y$ have the form
\begin{eqnarray}
W_x=W_x(s)&=&(s+s^{-1})\exp[2\pi(M_++\ii p_+)]+2D_x,\label{Wx}\\
W_y=W_y(\st)&=&(\st+\st^{-1})\exp[2\pi(M_++\ii q_+)]+2D_y,\label{Wy}
\end{eqnarray}
where $s=\exp[2\pi(z-\ii y)]$ and $\st=\exp[2\pi(z+\ii x)]$,
and where $D_x$, $D_y$ are complex constants.
The real and imaginary parts of $D_x$ and $D_y$ are moduli; but they
are not independent, so do not provide all the moduli.

The Nahm transform maps walls to walls, although in general
the gauge group, the topological charges, and the number of Dirac
singularities change \cite{CW12, C14}. In our case, however, these
properties do not change: the Nahm transform of a smooth SU(2) wall
of charge $(1,1)$ is again of that type.  The action of a Nahm transform
on the parameters and the moduli is as follows:
\begin{eqnarray}
(M_+,p_+,q_+) &\mapsto& (-M_+,-p_+,-q_+),\label{NahmParam}\\
D_x &\mapsto& -D_x\exp[-2\pi(M_++\ii p_+)],\label{NahmModuli}\\
D_y&\mapsto&-D_y\exp[-2\pi(M_++\ii q_+)]).
\end{eqnarray}
These expressions follow from the fact that the $x$-spectral curve,
given by $t^2-tW_x(s)+1=0$, is invariant under the Nahm transform,
which acts by interchanging the variables $t$ and $s$; and similarly for
the $y$-spectral curve \cite{CW12}.


\section{The asymptotic region of $\cM$}
In order to understand the role played by the parameters and the moduli,
let us first look at the asymptotic region of moduli space $\cM$, which consists
of those fields for which $|\Phi|_{z=0}\gg1$. It follows from this condition
that $D_x$ and $D_y$ have the approximate form
\begin{equation} \label{AsymptoticDxy}
 D_x\approx\cosh[2\pi(M+\ii p)], \quad D_y\approx\cosh[2\pi(M+\ii q)],
\end{equation}
with $M\gg\max\{1,M_+\}$.  Three of the four asymptotic moduli are
$M$ and $p,q\in(-\half,\half]$. The walls are located at values of $z$
for which $W_x(s)$ has zeros, and we see from (\ref{Wx}) that 
this occurs for $z=z_{\pm}=\pm(M-M_+)$; so we have two well-separated
walls. Note that $|D_x|\approx|D_y|$ up to exponentially small corrections,
so we could equally well have used the zeros of $W_y(\st)$ to define
the wall locations; but this is only true asymptotically, and not in the core
region of $\cM$. Each wall has a monopole embedded in it, the monopole
locations $\vR_{\pm}=(x_{\pm},y_{\pm},z_{\pm})$ being defined to be where
$W_x(s)=0=W_y(\st)$. Numerical solutions indicate that this is where
$\Phi$ is zero, and also where the energy density is peaked.
It follows from (\ref{Wx}, \ref{Wy}) that the location of the $z>0$
monopole is $\vR_+=(\half+q-q_+,\half-p+p_+,M-M_+)$.
\begin{figure}[htb]
\begin{center}
\includegraphics[scale=0.4]{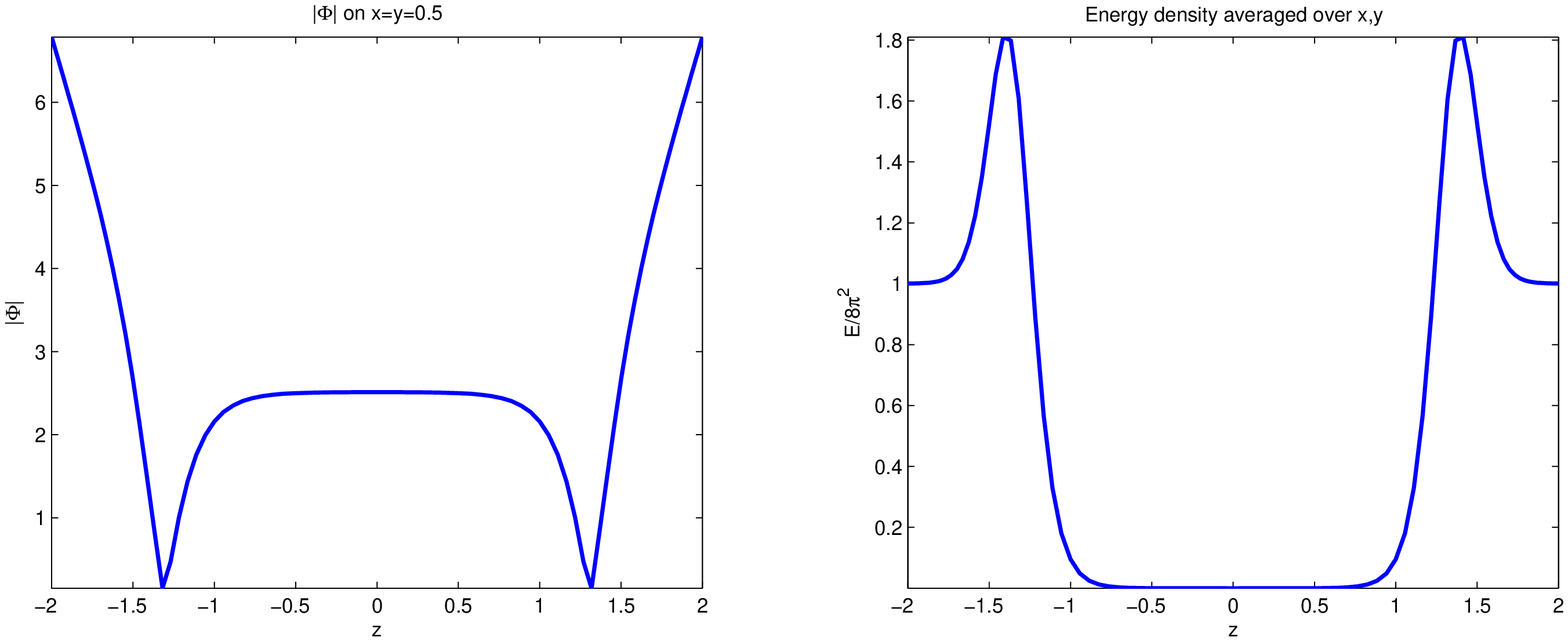}
\caption{Higgs field and energy density of a well-separated two-wall solution
 \label{Fig1}}
\end{center}
\end{figure}

The energy density is approximately zero for $z_-<z<z_+$ (between the two
walls), and tends to $8\pi^2$ as $z\to\pm\infty$. See Fig~1, which depicts
a solution with $M_+=-0.92$ and $D_x=D_y=6.21$; this solution was
obtained numerically by minimizing the functional (\ref{Energy}).
The left-hand plot is of $|\Phi|$ on the line $x=y=\half$, where the monopoles
are located. The right-hand plot is of the normalized, $xy$-averaged energy
density $(8\pi^2)^{-1}\int\cE\,dx\,dy$, as a function of $z$.
Between the walls, the function $|\Phi|$ is approximately constant;
in fact $|\Phi|\approx2\pi M$.

In view of the shape of the energy density, one might have
expected that $E_L$ could be reduced by moving the walls further apart,
{\sl ie.}\ by increasing $M$: it looks like an increase
$\delta M$ in $M$ would give $\delta E_L=-16\pi^2\,\delta M$, as
the central region (where $\cE$ is zero) increases in size. But in fact
as $M$ increases and the walls move apart, the energy contained
in each monopole increases by $8\pi^2\,\delta M$. This is because
each monopole resembles an $\RR^3$ monopole
with $|\Phi|_{\infty}=2\pi M$ and therefore energy $8\pi^2M$.
So the total energy $E_L$ is independent of $M$, as it must be from
(\ref{Bog}). Note, however, that stability
involves fixing the value of the parameter $M_+$, and reducing $M_+$ really
does lower the energy. This is analogous to having to fix the
boundary value of $|\Phi|$ in the $\RR^3$ case.

Furthermore, the size of each monopole core is proportional to $M^{-1}$, 
and therefore one may think of them as small SU(2) monopoles embedded in
an ambient U(1) field. So the asymptotic moduli are analogous to those of
the $\RR^3$ case: three moduli $(M,p,q)$ determine the relative location
of the two monopoles, and the fourth is a relative phase
$\omega\in(-\pi,\pi]$ between them. The asymptotic metric, in our
coordinates $(M,p,q,\omega)$, takes the hyperk\"ahler form \cite{HKM14}
\begin{equation} \label{AsympMetric}
 ds^2=\pi W(dM^2+dp^2+dq^2)+\pi W^{-1}[d\omega-8\pi(q\,dp-p\,dq)]^2,
\end{equation}
where $W=W(M)=8\pi(2M-M_+)$. Here, for simplicity, we have set
$p_+=q_+=0$.
Note from (\ref{AsympMetric}) that $R=M^{3/2}$ is an affine parameter
on asymptotic `radial' geodesics $p,q,\omega$ constant. The volume
${\rm Vol}_R$ of a ball of radius $R$ scales like ${\rm Vol}_R\sim R^{4/3}$,
and so $\cM$ is of ALH type \cite{C11}.


\section{The interior of $\cM$}
If $M_+\gg1$, then the monopoles are always well-localized:
the monopole size is small compared to unity even when
the walls are close together. The energy density is strongly peaked at the
locations of the two monopoles, one in each wall; if the monopoles coincide,
the energy is peaked on a well-localized torus.
So we expect that for $M_+\gg1$, we can interpret the moduli space
in terms of the locations and relative phase of the two monopoles,
taking account of the periodicity in the $x$- and $y$-directions.
If $M_+\ll-1$, the moduli space should
be the same (via the Nahm transform), although the corresponding
monopole picture will differ; in particular, the monopoles in this case
will not be well-localized when the walls are close together.

For the case when $M_+$ is close to zero,
one may also get information by looking at a neighbourhood
of the one explicit solution which is known, namely the
constant-energy solution. In a non-periodic gauge, this is
\begin{equation} \label{HomogSoln}
  \Phi_{(0)}=2\pi\ii z\,\sigma_3, \quad A_{(0)j}=\pi\ii(y,-x,0)\,\sigma_3.
\end{equation}
It has parameters $M_+=p_+=q_+=0$ and moduli
$D_x=D_y=0$, and its energy density has the constant value $8\pi^2$.
To understand nearby solutions, we examine
perturbations of (\ref{HomogSoln}); details have appeared in \cite{CW12},
and we summarize them here in a slightly different form.

If $\ve$ is an infinitesimal parameter, take the Higgs
field to be $\Phi=\Phi_{(0)}+\ve\Phi_{(1)}+\ve^2\Phi_{(2)}$, and similarly
for the gauge potential.
The equations for the first-order perturbation $(\Phi_{(1)},A_{(1)j})$
can be solved explicitly in terms of theta-functions.
If we write $\zeta=x+\ii y$, and define matrices $\Xi$ and $\Psi$ by
$2\Xi=A_{(1)x}+\ii\,A_{(1)y}$ and $2\Psi=A_{(1)z}+\ii\,\Phi_{(1)}$, then
the relevant solution is
\begin{equation} \label{PsiXi}
  \Psi=\ii\, g(\zeta) \bar{E} \,\sigma_+,
   \quad \Xi=\ii\, f(\bar{\zeta}) E \,\sigma_-,
\end{equation}
where $E=\exp(-2\pi z^2-2\pi\ii\bar{\zeta}y)$,
$2\sigma_{\pm}=\sigma_1\pm\ii\sigma_2$, and $f(\bar{\zeta})$,
$g(\zeta)$ are given by
\begin{equation}
 \overline{f(\bar{\zeta})}=\overline{C}_1[\vartheta_3(\pi\zeta)]^2 +
      \overline{C}_2[\vartheta_1(\pi\zeta)]^2,
\quad g(\zeta)=C_3[\vartheta_3(\pi\zeta)]^2+C_4[\vartheta_1(\pi\zeta)]^2.
\end{equation}
Here the $C_{\alpha}$ are complex constants, and we are using
standard theta-function conventions \cite{NIST}, with the nome of the
theta functions being $q=\ee^{-\pi}$.

Next, we obtain $\Phi_{(2)}$ etc by solving to second order in $\ve$.
This gives $\Phi_{(2)}=\ii\,\phi\,\sigma_3$ and $A_{(2)j}=\ii\, a_j\, \sigma_3$,
where $\phi$ and $a_j$ satisfy
\begin{equation}\label{poisson}
 \partial_j \phi + \ve_{jkl}\,\partial_k a_l=
  2\left(2\Re(f\bar{g}),\,2\Im(f\bar{g}),\,|g|^2-|f|^2\right)\exp(-4\pi z^2 -4\pi y^2).
\end{equation}
Here $f$ denotes $f(\bar{\zeta})$ and $g$ denotes $g(\zeta)$.
(Note that the coefficients in (\ref{poisson})
differ slightly from those in \cite{CW12}.)
The values of the parameters $(M_+,p_+,q_+)$ for the deformed solution
can be computed directly, and one gets
\begin{equation}\label{PerturbedParameters}
 M_+=\frac{\ve^2\Upsilon}{4\pi}\left(|C_3|^2+|C_4|^2-|C_1|^2-|C_2|^2\right),
  \quad
 p_+ + \ii q_+ = \frac{\ii\ve^2\Upsilon}{2\pi}\left(C_1C_3+C_2C_4\right),
\end{equation}
where
$\Upsilon=\int|\vartheta_1(\pi\zeta)|^4\exp(-4\pi y^2)\,dx\,dy\approx0.5902$.

Thus of the eight real quantities $C_{\alpha}$, three serve to set the parameters,
four are moduli, and the remaining one is gauge-removable, since
\begin{equation}\label{GaugeC}
  C_1\mapsto\ee^{\ii\theta}C_1, \quad C_2\mapsto\ee^{\ii\theta}C_2,\quad
  C_3\mapsto\ee^{-\ii\theta}C_3, \quad C_4\mapsto\ee^{-\ii\theta}C_4
\end{equation}
amounts to a gauge transformation. (This gauge freedom corresponds to
isorotation about the $\sigma_3$-axis, which leaves the field
(\ref{HomogSoln}) unchanged.)
To get the parameter values $M_+=p_+=q_+=0$, one may take
$C_3=C_2$ and $C_4=-C_1$; and the residual gauge freedom is 
$C_{\alpha}\mapsto-C_{\alpha}$. So for these parameter values,
the moduli space has a conical singularity at the point
(\ref{HomogSoln}): the ``tangent space'' there is $\RR^4/\ZZ_2$.
For $M_+\neq0$, however, the moduli space is smooth.

The expressions above enable us to describe the solutions which are
close to the constant-energy field (\ref{HomogSoln}), either directly
for small $\ve$, or by using them as starting configurations and then
minimizing the energy $E_L$ to get a numerical solution. This leads to the
following picture. If $C_1=C_2=0$, but $|C_3|^2+|C_4|^2\neq0$ and hence
$M_+>0$, one gets monopoles in the plane $z=0$.
In other words, $\Phi$ has a pair of zeros, which may coincide, on $z=0$;
and the energy density is peaked at those zeros as usual.
\begin{figure}[htb]
\begin{center}
\includegraphics[scale=0.8]{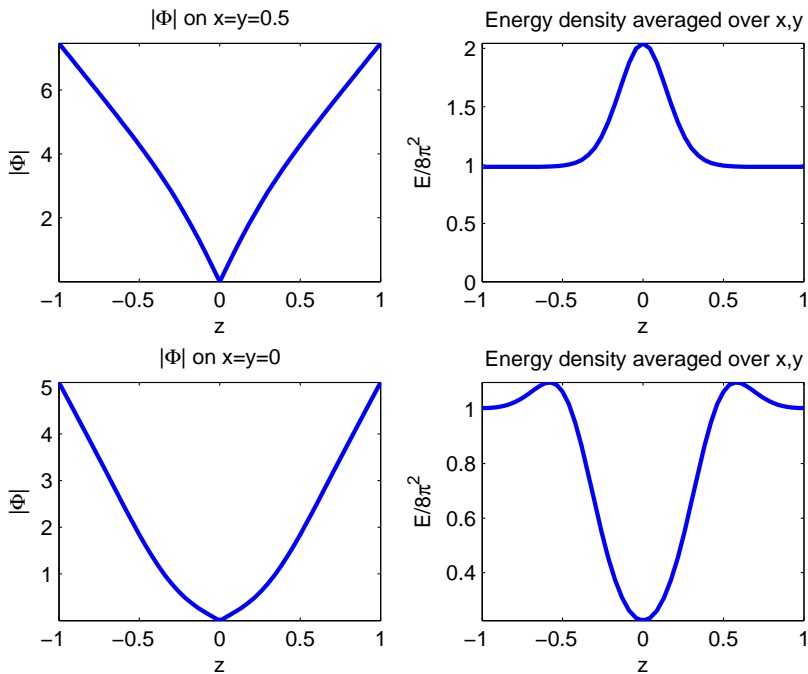}
\caption{Higgs field and energy density of two double-wall solutions
  \label{Fig2}}
\end{center}
\end{figure}
The top row of Fig~2 illustrates a numerically-generated solution which
is a non-infinitesimal version of the $C_{\alpha}=(0,0,1,0)$ case.
The quantities plotted are the same as in Fig~1. The solution has
$M_+=0.2$ and $D_x=D_y=1.8$.
There is a double monopole (a torus with its axis in the $z$-direction)
at $(x,y,z)=(\half,\half,0)$, and this is where the energy density is peaked.

If, however, $C_3=C_4=0$, but $|C_1|^2+|C_2|^2\neq0$ and hence
$M_+<0$, then $\Phi$ is identically zero on $z=0$, whereas the energy density
is minimal on $z=0$ and peaked off $z=0$. The bottom row of
Fig~2 depicts a non-infinitesimal version of the $C_{\alpha}=(0,1,0,0)$ case,
a solution having $M_+=-0.2$ and $D_x=D_y=-0.5$. 
The two solutions depicted in Fig~2 are Nahm transforms of each other,
with their parameters and moduli being related
as in (\ref{NahmParam}, \ref{NahmModuli}).


\section{Geodesic surfaces, geodesics, and trajectories}
One can identify several geodesics in $\cM$ as fixed-point sets of discrete
isometries, and this section describes a few of them, together with their
interpretation as monopole-scattering trajectories.
When the monopoles are well-localized, one may visualize such
isometries in terms of their action on the two-monopole system
viewed as a single rigid body, with three principal axes of inertia,
as in the $\RR^3$ case \cite{MS04}. The line joining the two monopoles
is called the (body-fixed) 3-axis, a head-on
collision results in a torus whose axis is the 1-axis, and the 2-axis
is the line along which the monopoles emerge after scattering.

Let $\tau_0$ denote rotation by $180^\circ$ in the $xy$-plane:
in other words $\tau_0:(x,y)\mapsto(1-x,1-y)$. Then $\tau_0$ maps
$(M_+,p_+,q_+)$ to $(M_+,-p_+,-q_+)$; so if we take $p_+=q_+=0$,
as we shall do from now on, then $\tau_0$ is a symmetry of the system,
preserving both the Bogomolny equation and the boundary conditions.
Also, $\tau_0$ leaves the relative phase $\omega$ of two well-separated
monopoles unchanged, and maps $(D_x,D_y)$ to $(\bar{D}_x, \bar{D}_y)$.
It follows that the fixed-point set of $\tau_0$
is a 2-dimensional geodesic surface $\cS$ in the moduli space $\cM$.

The quantities $D_x$ and $D_y$ are real-valued on $\cS$, and 
in the asymptotic region of the moduli space we have
$|D_x|\approx|D_y|\gg1$. So $\cS$ has four asymptotic components,
according to whether each of $D_x$ and $D_y$ is positive or
negative. This corresponds to having two monopoles, well-separated in
the $z$-direction, with the same $(x,y)$-location: namely one
of the four possibilities $(0,0)$, $(\half,0)$, $(0,\half)$ or $(\half,\half)$.
The 3-axis is in the $z$-direction, and the direction of the 1-axis in
the $xy$-plane corresponds to the relative phase $\omega$, which
is unrestricted. So each of the four asymptotic components is a cylinder,
on which the coordinates are  $M\gg1$ and $\omega\in S^1$.

In order for a monopole pair to be invariant under $\tau_0$, its 1-axis must
either be orthogonal to the $z$-axis (as in the asymptotic situation of the
previous paragraph) or parallel to it; this
gives two disjoint components of $\cS$, namely $\cS_1$ and $\cS_0$
respectively. (The same sort of thing happens in the singly-periodic
monopole-chain case \cite{MW13}: in that case, $\cM$ contains
a surface for which the 1-axis is orthogonal to the periodic axis, plus
two surfaces, isometric to each other, for which the 1-axis is
along the periodic axis.) As we shall see below, the four asymptotic
cylinders of $\cS$ referred to above are the ends of the single
component $\cS_1$.

We now find geodesics in $\cS_1$ and $\cS_0$ by imposing additional
symmetries. Two such isometries of $\cM$ correspond to reflections in
the $xy$-plane, namely
\begin{eqnarray}
  \tau_1 &:& x\mapsto1-x, \Phi\mapsto-\Phi, \\
  \tau_2 &:& x\mapsto y, y\mapsto x, \Phi\mapsto-\Phi.
\end{eqnarray}
Note that, on $\cS$, $\tau_1$ is equivalent to the reflection $y\mapsto1-y$,
and $\tau_2$ is equivalent to $x\mapsto-y$, $y\mapsto-x$; so it is
unnecessary to consider these reflections as well. In the asymptotic
region, requiring invariance under $\tau_1$ or $\tau_2$ has the effect
of restricting the direction of the 1-axis (the relative phase of the two
monopoles), and gives us geodesics in $\cS_1$.
The $\tau_1$-invariant fields have their 1-axis in the $x$- or $y$-direction,
while the $\tau_2$-invariant fields have their 1-axis along either $x=y$ or
$x=-y$. So in each asymptotic cylinder of $\cS_1$, we can identify four
geodesics, and each of them can be traced as it passes through the interior
of $\cS_1$, using the analogous $\RR^3$ scattering behaviour.
(Here we are imagining that the monopoles remain
well-localized throughout, which is the case if $M_+\gg1$.
In the $M_+\ll-1$ case, the moduli space and its geodesics are the same,
via the Nahm transform, but the scattering interpretation is necessarily
different.) For example, start on the asymptotic cylinder $D_x\approx D_y<0$
(monopoles on $x=y=0$), with the 1-axis in the $x$-direction. Then the two
incoming monopoles merge at $x=y=z=0$, separate along the $y$-axis,
re-merge at $(x,y,z)=(0,\half,0)$, separate in the $z$-direction, and finally
emerge in the asymptotic cylinder with $D_x>0$, $D_y<0$. Each pair of
asymptotic cylinders is connected by a geodesic (either $\tau_1$- or
$\tau_2$-invariant) in this way, and so they are the ends of the single
component $\cS_1$ of the surface $\cS$, as mentioned previously.

The fate of generic geodesics starting in the asymptotic region of $\cS_1$
is less clear, but it seems likely that (unlike in the example above)
they never emerge: they get trapped in the central region of $\cS_1$,
and continue travelling around the $z=0$ torus.

Let us now turn to geodesics on the other component of $\cS$, namely
$\cS_0$. As before, we first focus on the $M_+>0$ case, where the monopoles
are localized. They are necessarly confined to the $z=0$ plane
--- the two walls coincide, and the monopole motion takes place entirely
within this double wall. We can get a good picture by thinking of
perturbations of the constant-energy solution, as described in the
previous section. In particular, we take the subclass of perturbations
given by $C_1=C_2=0$: these fields are invariant under the $180^{\circ}$
rotation $\tau_0$, and in effect give us the surface $\cS_0$. We
fix $|C_3|^2+|C_4|^2$ in order to fix $M_+>0$, and factor out by the phase
(\ref{GaugeC}), so $\cS_0$ is a 2-sphere $S^2$ on which
$\xi=C_4/C_3$ is a stereographic coordinate. Note, however, that
the metric on $\cS_0$ is not the standard 2-sphere metric.

Four points on this sphere, namely $\xi=0,\infty, 1 \mbox{ and }-1$,
correspond to toroidal double-monopoles at
$(x,y)=(\half,\half),(0,0),(0,\half)\mbox{ and }(\half,0)$ respectively.
The point $\xi=\ii$ corresponds to a pair of monopoles at
$(x,y)=(\quar,\quar)\mbox{ and }(\thrquar,\thrquar)$, while $\xi=-\ii$
corresponds to a pair of monopoles at
$(x,y)=(\quar,\thrquar)\mbox{ and }(\thrquar,\quar)$.
Imposing various additional symmetries then gives closed geodesics
on $\cS_0$. For example,
invariance under $\tau_1$ gives a geodesic passing through
$\xi=0,1,\infty\mbox{ and }-1$ in that order; whereas
$\tau_2$-invariance gives a geodesic passing through
$\xi=0,\ii,\infty\mbox{ and }-\ii$. These correspond to closed
trajectories in which the two monopoles repeatedly scatter
at right angles within the periodic $xy$-plane, via the toroidal
double-monopoles listed above.

\begin{figure}[htb]
\begin{center}
\includegraphics[scale=0.55]{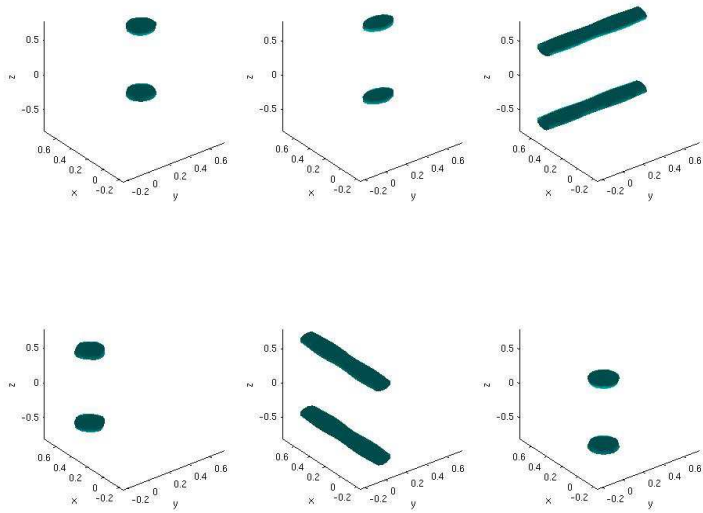}
\caption{Part of a closed trajectory on $\cS_0$ with $M_+<0$.
  \label{Fig3}}
\end{center}
\end{figure}
All this has a Nahm-transformed counterpart, with $M_+$ negative but
close to zero. The fields are perturbations of the constant-energy
solution (\ref{HomogSoln}) with $C_3=C_4=0$.
Recall that the Higgs field is now identically
zero on $z=0$, and that the energy density $\cE$ is peaked off $z=0$.
A geodesic can be visualized in terms of the
movement of these energy peaks, and one such closed trajectory
(or rather half of it) is illustrated in Fig~3. This shows six solutions,
corresponding to six points on the curve
$(C_1,C_2)=(\cos\eta,\sin\eta)$ for $0\leq\eta\leq\pi$, which is a closed
geodesic in $\cS_0$. Each of the figures is a
3-dimensional plot of the surface $\cE(x,y,z)=0.95\max(\cE)$,
and so it indicates where the energy density is peaked.
The upper-left figure shows peaks on $(x,y)=(\half,\half)$.
These elongate in the $y$-direction (top row), and re-localize as peaks
on $(x,y)=(\half,0)$ (the lower-left figure). They then elongate in the
$x$-direction before re-forming as peaks on $(x,y)=(0,0)$. The rest
of the closed trajectory (not shown) then proceeds via peaks at
$(x,y)=(0,\half)$ before returning to the initial field.


\section{Concluding remarks}
In this paper, we have studied doubly-periodic BPS 2-monopole solutions,
or double monopole walls. The moduli space of centred 2-monopole fields
is a 4-dimensional manifold $\cM$, and the moduli can be interpreted
in terms of the relative monopole positions and phases.
Even though the metric of $\cM$ is not known explicitly (except
in its asymptotic region), geodesics can be identified as fixed-point sets
of discrete isometries, and these may be interpreted as the interaction of
parallel monopole walls, or of the monopoles embedded in the walls.

For the gauge group SU(2), there are two topological charges $(Q_-,Q_+)$,
and the number of monopoles is $N=Q_-+Q_+$. In this paper, we have
only dealt with the
charge $(1,1)$ case. For walls of charge $(0,2)$ or $(2,0)$, many of the
details are similar, in particular the geometry of the moduli space.
Rather less is currently known about $N>2$ solutions, and it would be
interesting to investigate the existence of highly-symmetric
multi-monopole-wall configurations along similar lines to the $\RR^3$
case \cite{MS04}.

It would also be interesting to extend the analysis to the case of walls
which have hexagonal rather than square symmetry. In particular, this
would be relevant to the closely-related topic of monopole bags in $\RR^3$
\cite{B06, LW09, H12, HPS12, T13}, which have curved hexagonal
monopole walls separating their interior and exterior regions. It
also motivates the question of the general dynamical behaviour of
monopole walls, where double periodicity is not necessarily maintained,
and so there are infinitely many degrees of freedom; little is currently
known about this more general situation.


\bigskip\noindent{\bf Acknowledgments.}
Both authors were supported by the UK Particle Science and Technology
Facilities Council. For RSW this was through the Consolidated Grant No.\ 
ST/J000426/1.


\end{document}